\documentstyle[12pt]{article}
\begin{document}

\def\beq{\begin{equation}}
\def\eeq{\end{equation}}
\def\bce{\begin{center}}
\def\ece{\end{center}}
\def\ben{\begin{enumerate}}
\def\een{\end{enumerate}}
\def\ul{\underline}
\def\ni{\noindent}
\def\nn{\nonumber}
\def\bs{\bigskip}
\def\ms{\medskip}
\def\wt{\widetilde}
\def\wh{\widehat}
\def\Tr{\mbox{Tr}\ }

\hfill DFTUZ 95-01

\hfill HUPD  9505

\hfill January, 1995

\vspace*{3mm}

\begin{center}

{\LARGE \bf One loop renormalization of
the four-dimensional theory for quantum dilaton gravity}

\vspace{15mm}

\medskip

{\sc Ilya L. Shapiro} \\
Departamento de Fisica Teorica, Universidad de Zaragoza,
50009, Zaragoza, Spain.
\footnote{On leave from Tomsk Pedagogical
Institute, 634041 Tomsk, Russia. \\
E-mail: shapiro@dftuz.unizar.es}
\vskip 10mm
{\sc Hiroyuki Takata}\\
  Department of Physics,
 Hiroshima University,  Higashi-Hiroshima 724,
Japan. \footnote{ E-mail:
 takata@theo.phys.sci.hiroshima-u.ac.jp}\\
\vskip 5mm
\vspace{15mm}

{\bf Abstract}

\end{center}

We study the one loop renormalization in the most general metric-dilaton
theory with the second derivative terms only. Classical action includes
three arbitrary functions of dilaton. The general theory can be divided
into two classes, models of one are equivalent to conformally coupled
with gravity scalar field and also to general relativity with cosmological
term. The models of second class have one extra degree of freedom which
corresponds to dilaton. We calculate the one loop divergences for the
models of second class and find that the theory is not renormalizable off
mass shell. At the same time the  arbitrary functions of dilaton in the
starting action can  be fine-tuned in such a manner that all the higher
derivative counterterms disappear on shell. The only structures in both
classical action and counterterms, which survive on shell, are the
potential (cosmological) ones. They can be removed by renormalization of
the dilaton field which acquire the nontrivial anomalous dimension, that
leads to the effective running of the cosmological constant. For some of
the renormalizable solutions of the theory the observable low energy
value of the cosmological constant is small as compared with the
Newtonian constant. We also discuss another application of our result.
In particular, our calculations in a general dilaton model in original
variables give the possibility to estimate quantum effects in
$\Lambda+\alpha R+\beta R^2$ theory.

\vspace{4mm}

\newpage

\section{Introduction}

Recently it has been a considerable interest to the metric - scalar  gravity in
four dimensions. The active research in this field was inspired by
different reasons (see \cite{DEF} for the interesting discussion of
the subject). In particular,
the effective action of (super)string depends on both metric and dilaton
(see, for example, \cite{GSW}).
Such an effective action arise in a form of the power series in a string
loop parameter $\alpha'$, and the standard point of view is that
the higher orders in such an expansion correspond to higher energies.
 From this point of view at lower energy scale the action for gravity
has the form of the lower derivative dilaton action. From another hand
the presence of dilaton in a low derivative gravity action leads to the
inflationary cosmological solution, that enables one to solve some
specific problems in the field of cosmology. The problem of classical
solutions and the cosmological phase transitions in a dilaton theory has
been extensively studied  (see, for example,
\cite{bar2,maeda,wein,CO,DN}).
Moreover it turned out that some special version of the dilaton gravity
is classically equivalent to the restricted higher derivative gravity theory
which include only square of scalar curvature in addition to Hilbert-Einstein
action \cite{whit,bar2,maso} (see also the last paper for more complete
references).
This theory is also of big cosmological
interest because it enables one to construct the inflationary solutions
\cite{star,baot,MMS,hans,CO}).

Perhaps the completely consistent theory of quantum gravity can be
constructed within the string model, and gravity will be described by
effective action within this frame. However the string theory can be valid
at the Planck energies and above, and if one wish to deal with the
energies below   Planck scale, it is natural to suppose that the quantum
effects of gravity will be related with some low energy action.
One can, for instance, apply higher derivative gravity for this purposes.
Higher derivative gravity is renormalizable
\cite{9,vortyu} and allow the renormalization group study of some physical
effects like asymptotic freedom \cite{11,12,bksvw}
and phase transitions \cite{bosh,odsh1},
but not unitary (at least within the
usual perturbation scheme
(see \cite{book} for the introduction and more complete
references). Thus at the moment
we do not have any consistent theory which is applicable below Planck scale
and any researchment in this field is based on the choice of some model,
which allow us to explore some quantum gravity effects.
In present paper we consider the four dimensional metric-dilaton
model including the second derivative terms only. We choose the most general
action including arbitrary functions $A, B, C$ of the dilaton $\phi$.
\beq
S= \int d^4x \sqrt{-g}\; \{ A(\phi)g^{\mu\nu}\partial_{\mu}\phi
\partial_{\nu}\phi + B(\phi)R + C(\phi) \}
                                                    \label{0.1}
\eeq
that covers all special cases including the string inspired action,
special (relevant from cosmological viewpoint) case of
higher derivative gravity and also admit some other interesting applications.
Such a model is non-renormalizable that can be seen already from power
counting consideration.
Indeed one can suppose that all necessary counterterms are
introduced from the very beginning, but then the finite parts of the amplitudes
and also the "beta functions" for the generalized couplings
will be ill defined because of relevant
gauge fixing and parametrization dependence and therefore any
analysis becomes inconsistent.
However there are a few possibilities to obtain some sound results for the
theory (\ref{0.1}) on quantum level. First of all there is some interest to
explore the one-loop renormalization of the theory and to compare the
results with the ones for General Relativity \cite{hove}.
In the last case all the
one-loop counterterms vanish (if the cosmological term is lacking)
on mass shell and hence the one-loop $S$-matrix
is finite. The theory with cosmological constant is renormalizable
\cite{cosm}, however if one introduce the matter fields the one loop
renormalizability is lacking even on mass shell.
It should be interesting to know, whether it is so for the dilaton
model (\ref{0.1}). In this case the situation is much
more complicated, because the amount of possible counterterms is
essentially higher as compared with the pure metric theory. It turns out,
however, that it is possible to reduce the counterterms to the few
structures which survive on mass shell.

Furthermore, if the consideration is restricted by
the one-loop on shell case, then the theory with cosmological term $C(\phi)$
can be renormalizable that leads to some general conjectures about the
high energy behavior of quantum gravity \cite{11}.
Next, we can restrict ourselves by some special backgrounds where the
theory is renormalizable. For example, the cosmological inflationary
background provides
the renormalizability of the special higher derivative model which is
the particular case of the above model \cite{OV}. On the other hand one
can introduce an additional constraint on the background dilaton and
regard it as constant. This way is also of some cosmological interest,
because the renormalizability in the potential sector enables one to
evaluate the significance of quantum gravity for the cosmological phase
transitions.

The action (\ref{0.1}) may be viewed as the second derivative part of the
general (fourth derivative) model of the dilaton gravity, which has been
recently inversigated in \cite{ejos}.
In \cite{ejos} we have restricted ourselves by the case when only
 the scalar field is the quantum variable.
Despite the general case is very interesting, the
explicit calculations are too cumbersome  because of the presence of
higher derivatives. Here we perform the one-loop calculations in the theory
(\ref{0.1}), considering both fields $\phi$ and $g_{\mu\nu}$ as quantum ones.
We start with the general model (\ref{0.1}) and then turn to the analysis of
special cases.

The paper is organized as follows. In section 2 we discuss the different
conformally equivalent forms of the theory (\ref{0.1}), and show
that all of them can be divided into two sets.  Models of one set are
classically equivalent to conformal scalar - metric theory and,
simultaneously, to
General Relativity. The models of second set include physical
degrees of freedom, corresponding to the dilaton (or conformal factor),
and in forthcoming
sections we restrict the consideration only by the models of this class.
In section 3 the general
structure of renormalization of (\ref{0.1}) is explored both off and on
mass shell.
In section 4 we calculate of the one-loop counter terms.
To make this we apply the method which was developed in \cite{odsh}
within the two dimensional dilaton gravity. It turns out that it is
useful in $d=4$ as well, and not only in the model (\ref{0.1}) but also in the
higher derivative dilaton gravity formulated recently in
\cite{ejos} (see also discussion in \cite{sh94}).
In section 5 the concrete analysis of the on shell renormalization
of the model is performed. Here we fine tune the functions $A$ and $B$ to
provide one-loop finiteness of the theory without $C$ term.
If the potential term is included then the one loop on shell
renormalizability require the vanishing of the Einstein counterterm.
As a result we face with the cosmological type divergences only,
and it turns out that they can be removed by renormalization of the
scalar dilaton field.
  In section 6 the renormalization of the dilaton theory (\ref{0.1}) interacted
with  matter fields is discussed. It turns out that qualitatively the
structure of counterterms is the same as in the Einstein gravity, and
the dilaton - metric theory with matter is non-renormalizable even
on mass shell.
In section 7 we give the qualitative discussion of the renormalization in
two special cases, one of them is rather interesting and has to be analyzed
separately.
The last  section consists in discussion of the results.

\section{General notes on the dilaton gravity}

If we are interested to understand the parametrization dependence of the
dilaton action, it is useful to start with the simple particular case
of the general action (\ref{0.1}).
\beq
S= \int d^4x \sqrt{-g'} \; \{ R'\Phi+ V(\Phi) \}
                                                    \label{1.1}
\eeq
Here the curvature $R'$ corresponds to the metric $g'_{\mu\nu}$ and
$g'= \det (g'_{\mu\nu})$. Let now transform this action to new variables
$g_{\mu\nu}$ and $\phi$ according to
\beq
g'_{\mu\nu}=g_{\mu\nu}e^{2\sigma(\phi)},\;\;\;\;\;\;\;\;\;\;\;
\Phi=\Phi(\phi)                                     \label{1.2}
\eeq
where $\sigma(\phi)$ and $\Phi(\phi)$ are arbitrary functions of
$\phi$. In a new variables the action becomes:
\beq
S= \int d^4x \sqrt{-g}\; \{ \Phi(\phi)Re^{2\sigma(\phi)}+
6(\nabla\phi)^2e^{2\sigma(\phi)}[\Phi\sigma'+\Phi']\sigma'
+ V(\Phi(\phi))e^{4\sigma(\phi)} \}
                                                    \label{1.3}
\eeq
Therefore we are able to transform the particular action (\ref{1.1})
to the general form (\ref{0.1}) with
\beq
A(\phi)=6e^{2\sigma(\phi)}[\Phi\sigma'+\Phi']\sigma',\;\;\;\;\;\;\;\;\;\;\;
B(\phi)=\Phi(\phi) e^{2\sigma}                     \label{1.4}
\eeq
It is quite reasonable to explore the inverse problem, that is to
find the form of $\sigma(\phi)$ and $\Phi(\phi)$ that correspond to the
given $A(\phi)$ and $B(\phi)$. One can find that in this case $\sigma(\phi)$
and $\Phi(\phi)$ are  defined from the equations
\beq
 A=6B_1\sigma_1-6B(\sigma_1)^2,
\ \  \ \ \ \ \ \ \  \Phi = B e^{-2\sigma}                     \label{1.7}
\eeq
Substituting (\ref{1.7}) into (\ref{1.3}) we find that in a new variables the
action have the form
\beq
S= \int d^4x \sqrt{-g} \;\{ A(\phi)g^{\mu\nu}\partial_{\mu}\phi
\partial_{\nu}\phi + B(\phi)R + \left( \frac{B}{\Phi}\right)^2
V(\Phi(\phi))\}
                                                              \label{1.8}
\eeq
where the last term is nothing but $C(\phi)$ from(\ref{0.1}).

It is easy to see that the above transformations lead to some restrictions
on the functions $A(\phi)$ and $B(\phi)$. Let us consider the special case of
Einstein gravity that is to put $\Phi=const$. One can rewrite this condition
in terms of $A(\phi)$ and $B(\phi)$. Note that
\beq
2AB -3(B_1)^2 = -3\;(\frac{d\Phi}{d\phi})^2 \;e^{4\sigma(\phi)}   \label{1.5}
\eeq
Here and below the lower index show
the order of derivative with respect to $\phi$. For instance,
\beq
B_1 = \frac{dB}{d\phi},\ \ \ \ \ A_2  = \frac{d^2 B}{d\phi^2},\ \ \ \
\sigma_1 = \frac{d\sigma}{d\phi}, \ \ \ etc.
                                                              \label{1.6}
\eeq
Hence it is clear that the case of $\;2AB -3(B_1)^2 =0\;$ qualitatively differs
from another ones. Let us now comment this amusing case. We start with
the most simple example $A=1, B=\xi \phi^2$ where $\xi=\frac{1}{6}$.
Then the equation (\ref{1.7})
can be easily solved and we  obtain $\sigma(\phi) = \sigma_0+\ln|\phi|$
and $\Phi=\frac{1}{6}e^{2\sigma_0}=const$. Next, substituting these
expressions  into (\ref{1.8}) we find that in a new variables the transformed
action has the form
\beq
S= \int d^4x \sqrt{-g} \{ g^{\mu\nu}\partial_{\mu}\phi
\partial_{\nu}\phi + \frac{1}{6}R\phi^2 + \lambda \phi^4\}
                                                              \label{1.10}
\eeq
where $\lambda=e^{-4\sigma_0}\;V(\frac{1}{6}\;e^{2\sigma_0})$.
Thus we see that in a new variables the starting Hilbert-Einstein action
(\ref{1.1}) (remind that $\Phi$ is constant here)
corresponds to the conformally
coupled scalar field $\phi$. The extra scalar degree of freedom in (\ref{1.10})
is compensated by extra symmetry - local conformal invariance.
Both theories are
equivalent on the classical level. On quantum level the conformal invariance
of the theory (\ref{1.10}) will be probably broken because of the
non-invariance of the measure of path integral over the metric (see \cite{shja}
for the discussion of this point in Weyl gravity. Indeed it is
not completely sufficient, and the conformal version  should be
investigated separately). Thus the new anomalous degree of freedom
starts to propagate and the equivalence of two theories can be violated.
Let us notice that the same frame for Einstein gravity has
been recently used in \cite{25} for the investigation of  $2+\varepsilon$
quantum gravity.

And so, all the theories (\ref{0.1}) can be divided into two sets. First set
is labeled by $2AB -3(B_1)^2 =0$, it is conformally equivalent to the General
Relativity with cosmological constant. For the second set
$2AB -3(B_1)^2 \neq 0$. Such models are conformally equivalent to (\ref{1.1})
with non-constant $\Phi$. Below we shall deal only with the theories
of the second type. On classical level the change
 of dynamical variables can be compensated by the change of the functions
$A(\phi), B(\phi), C(\phi)$. However, as it was recently discussed by
Magnano and Sokolowski \cite{maso},
the natural choice of the frame is preferable from
physical point of view already on classical level. One can face the same
situation in the quantum theory as well. To see this, let us consider one
interesting particular case \cite{OV}.
If one put the potential term in (\ref{1.1})
in the special way and make the shift of the field $\Phi=\phi-\phi_0$,
where $\phi_0=const$ then the resulting theory
\beq
S= \int d^4x \sqrt{-g}\;\{\frac{1}{4\,\alpha}\phi^2+ R(\phi-\phi_0)+\Lambda \}
                                                    \label{1.12}
\eeq
is equivalent to the special version of higher derivative quantum gravity
\beq
S=\int d^4x\sqrt{-g}\{\;-\alpha R^2-\frac{1}{G}R+\Lambda\}\label{1.11}
\eeq

However on quantum level it is so only if we do not introduce into the
generating functional of the Green functions
 the external source for the auxiliary field $\phi$.
If one make some nonlinear change of variables like the conformal transform
described above, the auxiliary field and the conformal factor of the
metric are mixed and we likely lose the simple relation between
(\ref{1.11}) and (\ref{1.12}).

\section{The structure of the counterterms off and on mass shell}

The main purpose of the present paper is to investigate the theory
(\ref{0.1}) on quantum level within the one - loop approximation.
The simple consideration based on power counting
shows that the theory is non-renormalizable just as General Relativity.
At the same time
at one-loop order the last theory is renormalizable on mass shell
\cite{hove}. This
property holds even if the cosmological constant is included to the action
\cite{cosm} that enables one to apply some kind of renormalization group
approach for it's study  \cite{11}.
That is why  it looks interesting to consider the renormalization of
our theory on mass shell. The next reason to do this is the gauge and
parametrization independence  of the effective action on mass shell.

In this section we write down the classical equations of motion
and the possible divergent structures, taking into account only the
one loop order. Then we find some simple relations between counterterms
and consider the divergent structures which are possible on shell.
The equations of motion in the theory (\ref{0.1}) have the form
$$
 B R^{\mu\nu}+ g^{\mu\nu}
\left[\left( B_2 -\frac{A}{2} \right)(\nabla \phi)^2
-\frac{R+C}{2}+B_1(\Box \phi)\right]
+ (A -B_2)(\nabla^{\mu} \phi)(\nabla^{\nu}\phi)
 -B_1 (\nabla^{\mu} \nabla^{\nu} \phi) =0
$$
\beq
 B_1 R + C_1 -A_1 (\nabla \phi)^2 -2A (\Box \phi) =0     \label{2.1}
\eeq

Before going on to discuss the renormalization of the theory,
one have to define
the classical dimension of the field $\phi$. The form of the
starting Lagrangian shows that there is some dimensional parameter $M$
from the very beginning. One can introduce such dimensional parameter
in a different ways, that corresponds to different classical dimensions
of the scalar field $\phi$. For instance, in
the case of dimension-less $\phi$ the arbitrary functions $A, B, C$
include the dimensional parameter $M$ in a trivial way $A, B \sim M^2$
and $C \sim M^4$.
On the contrary, if the dimension of $\phi$ is chosen as unity,
then (if we want to consider arbitrary functions $A, B, C$),
they depend on the ratio $\frac{\phi}{M}$. Of course  the results of the
explicit
(one-loop in our case) calculations do not depend on this choice, and thus
we can regard the dimension of $\phi$ according to our convenience.
On this stage it is better to
consider the dimension-less $\phi$. Then the arbitrary functions $A,B,C$
do not contain the dimensional parameter $M$.
Another advantage of this choice is that $\phi$ and metric have
an equal dimensions and therefore the power counting in
a dilaton theory is essentially the same as compared with General Relativity.

If one is interested only in the one-loop divergences, then the counterterms
contains the covariant terms of fourth order in derivatives.
 The most general action of this type has the form \cite{ejos}:
$$
\Gamma_{div}^{1-loop}= - \frac{1}{16 \pi^2 (n-4)}
\int d^4x\sqrt{-g}
[ c_w C^2 + c_r R^2 + c_4 R(\nabla \phi)^2 + c_5 R(\Box \phi )+
 c_6 R^{\mu \nu}(\nabla_{\mu} \phi)(\nabla_\nu \phi) +
$$
\beq
+c_7 R +
c_8 (\nabla \phi)^4 + c_9 (\nabla \phi)^2(\Box \phi)+
 c_{10} (\Box \phi)^2 + c_{11}  (\nabla \phi)^2 +c_{12} ]+
(s.t.)                \label{2.2}
\eeq
where $C^2=C_{\mu\nu\alpha\beta }C^{\mu\nu\alpha\beta}$ is the square of Wyle
tensor and$(\nabla\phi)^2=g^{\mu\nu}\;\nabla_\mu\phi\;\nabla_\nu\phi$. $"s.t."$
means "surface terms". All $c_{w,r,4,...,12}$ are some
functions of $A(\phi), B(\phi), C(\phi)$ and their derivatives.
One can easily check the following reduction formulas which show the surface
form of the other possible structures \cite{ejos}.
$$
c_{13}(\nabla^\mu R)(\nabla_\mu\phi)=-c'_{13} R(\nabla_\mu\phi)^2-c_{13}R
(\Box\phi) + (s.t.)
$$
$$
c_{14}R_{\mu\nu}(\nabla^\mu\nabla^\nu\phi)=-c'_{14}R_{\mu\nu}
(\nabla^\mu\phi)(\nabla^\nu\phi)+{1\over 2}
c'_{14}R(\nabla\phi)^2+{1\over 2}c_{14}R(\Box\phi)  + (s.t.)
$$
$$
c_{15}(\nabla^\nu\phi)(\Box\nabla_\nu\phi)=-c'_{15}(\nabla\phi)^2
(\Box\phi)-c'_{15}(\Box\phi)^2+c'_{15}R_{\mu\nu}(\nabla^\mu\phi)
(\nabla^\nu\phi) + (s.t.)
$$
$$
c_{16}(\nabla_\nu\nabla_\mu\phi)^2={1\over 2}c''_{16}(\nabla\phi)^4+
{3\over 2}c'_{16}(\nabla\phi)^2(\Box\phi)+
c_{16}(\Box\phi)^2-c_{16}
R_{\mu\nu}(\nabla^\mu\phi)(\nabla^\nu\phi)  + (s.t.)
$$
$$
c_{17}(\nabla^\nu\Box\nabla_\nu\phi)=
c''_{17}(\nabla\phi)^2(\Box\phi)+
c'_{17}(\Box\phi)^2-c'_{17}R_{\mu\nu}(\nabla^\mu\phi)(\nabla^\nu\phi)
 + (s.t.)
$$
$$
c_{18}(\Box^2\phi)=c''_{18}(\nabla\phi)^2(\Box\phi)+
c'_{18}(\Box\phi)^2 + (s.t.)
$$
$$
c_{19}(\Box R)=c''_{19}R(\nabla\phi)^2+c'_{19}R(\Box\phi) + (s.t.)
$$
$$
c_{20}(\nabla_\nu\phi)(\nabla_\mu\phi)(\nabla^\nu\nabla^\mu\phi)=
-{1\over 2} c'_{20}(\nabla\phi)^4
-{1\over 2}c_{20}(\nabla\phi)^2(\Box\phi) + (s.t.)
$$
\beq
c_{21}(\nabla^\nu\phi)(\nabla_\nu\Box\phi)=-c'_{21}(\nabla\phi)^2
(\Box\phi)-c_{21}(\Box\phi)^2 + (s.t.)               \label{2.3}
\eeq
Here $c_{13,..,21} = c_{13,..,21}(\phi)$ are some (arbitrary) functions.

Thus the power counting consideration and the account of symmetries
show that the possible counterterms have complicated form and
differs from the classical action. Therefore the theory is expected to
be non-renormalizable off shell. Let us now discuss the
renormalization on mass shell. For this purpose we shall apply the equations
of motion (\ref{2.1}) and the reduction formulas (\ref{2.3}) and rewrite
the counterterms (\ref{2.2}) and the classical action (\ref{0.1}) in a
maximally simple form. In particular, from the equations of motion (\ref{2.1})
one can get the following relations
$$
(\nabla \phi)^2=x R + y,
\;\;\;\;\;\;\;\;\;\;\;\;\;\;\;\;\;
(\Box \phi)=z R + w
$$
\beq
(\nabla^{\mu} \nabla^{\nu} \phi) = r (\nabla^{\mu} \phi)(\nabla^{\nu}\phi)
   + s R^{\mu\nu} + t g^{\mu\nu} R + u g^{\mu\nu}         \label{2.4}
\eeq
where
$$
x(\phi)=
{{2\,A\,B - 3\,{{B_{1}}^2}}\over {-2\,{A^2} - 3\,A_{1}\,B_{1} + 6\,A\,B_{2}}}
$$$$
y(\phi)=
{{4\,A\,{\rm C} - 3\,B_{1}\,{\rm C}_{1}}\over
   {-2\,{A^2} - 3\,A_{1}\,B_{1} + 6\,A\,B_{2}}}
$$$$
z(\phi)=
{{-\left( B\,A_{1} \right)  - A\,B_{1} + 3\,B_{1}\,B_{2}}\over
   {-2\,{A^2} - 3\,A_{1}\,B_{1} + 6\,A\,B_{2}}}
$$$$
w(\phi)=
{{2\,{\rm C}\,A_{1} + A\,{\rm C}_{1} - 3\,B_{2}\,{\rm C}_{1}}\over
   {2\,{A^2} + 3\,A_{1}\,B_{1} - 6\,A\,B_{2}}}
$$$$
r(\phi)=
{{A - B_{2}}\over {B_{1}}}, \;\;\;\;\;\;\;\;\;\;\;\;\;\;\;\;\;\;\;\;\;
s(\phi)=
{B\over {B_{1}}}
$$
$$
t(\phi)=
{{B\,A_{1}\,B_{1} + A\,{{B_{1}}^2} - 2\,A\,B\,B_{2}}\over
   {2\,B_{1}\,\left( -2\,{A^2} - 3\,A_{1}\,B_{1} + 6\,A\,B_{2} \right) }}
$$
\beq
u(\phi)=
{{2\,{A^2}\,{\rm C} + {\rm C}\,A_{1}\,B_{1} - 2\,A\,{\rm C}\,B_{2} -
 A\,B_{1}\,{\rm C}_{1}}\over {2\,B_{1}\,\left( 2\,{A^2} +
3\,A_{1}\,B_{1} - 6\,A\,B_{2} \right) }}           \label{2.5}
\eeq
Next, combining the equations of motion (\ref{2.1}) and the reduction formula
for $c_{14}$ (\ref{2.3}) we find
$$
\int d^4 x \sqrt{-g}\;c_6 R_{\mu \nu} (\nabla^{\mu}\phi)(\nabla^{\nu}\phi)
= \int d^4 x \sqrt{-g}\; [\;
\frac{1}{2}c_6 x R^2 + \frac{1}{2}c_6 R+
$$
\beq
+f(\phi)\; \left\{
 -s R_{\mu \nu}^2 + \left( \frac{1}{2} r x + \frac{1}{2} z -t\right)R^2
+\left(\frac{1}{2}ry+\frac{1}{2}w-u\right) R\right\} ]    \label{2.6}
\eeq
where $f(\phi)$ is solution of the following differential equation
\beq
f_1(\phi) + f(\phi) r(\phi) - c_6(\phi) = 0                  \label{2.7}
\eeq
Since the last equation have solution for any $r(\phi)$ and $c_6(\phi)$,
we find that the on shell one loop divergences for our dilaton model
(\ref{0.1}) can be reduced to  the form of higher derivative terms without
explicit kinetic terms for the
dilaton. Of course one can choose another basis and express everything
in terms of dilaton structures only, removing the terms with higher
powers of curvature.

\section{One-loop calculations}

In this section we shall present the details of the calculation of the
one-loop counterterms of the theory (\ref{0.1}).
For the purpose of calculation of the divergences we will
apply the background field method and the Schwinger-De Witt technique.
The features of the metric-dilaton theory do not
lead to the necessity of some modifications of the calculational scheme,
basically developed in the similar two-dimensional theory \cite{odsh}.

Let us start with the usual splitting of the fields
into background $g_{\mu\nu}, \phi$
and quantum $h_{\mu\nu}, \varphi$ ones
\beq
\phi \rightarrow \phi' = \phi + \varphi\;,\;\;\;\;\;\;\;\;\;\;\;\;\;\;
\;   g_{\mu\nu} \rightarrow g'_{\mu\nu} + h_{\mu\nu}         \label{2.8}
\eeq
The one-loop
effective action is given by the standard general expression
\beq
\Gamma={i \over 2}\;\Tr\ln{\hat{H}}-i\;\Tr\ln {\hat{H}_{ghost}},\label{2.9}
\eeq
where $\hat{H}$ is the bilinear form of the action (\ref{0.1}) with
added gauge fixing term and $\hat{H}_{ghost}$
is the bilinear form of the gauge ghosts action.
To perform the calculations in a most simple way one needs to
introduce the special form of the gauge fixing term:
\beq
S_{gf} = \int d^4 x \sqrt{-g}\;\chi_{\mu}\;\frac{\alpha}{2}\;\chi^{\mu}
                                                           \label{2.10}
\eeq
where $\chi_{\mu} = \nabla_{\alpha} \bar{h}_{\mu}^{\,\alpha}+
\beta\nabla_{\mu}h+\gamma \nabla_{\mu} \varphi$, $h=h_{\mu}^{\mu},\;
\bar{h}_{\mu\nu}=h_{\mu\nu}-\frac{1}{4}\;hg_{\mu\nu}$ and $\alpha, \beta,
\gamma$ are some functions of the background dilaton, which can be tuned
for our purposes. For instance, if one choose these functions as
follows
\beq
\alpha=-B\;\;,\;\;\;\;\beta=-\frac{1}{4}\;\;,\;\;\;\;\gamma=-\frac{B_1}{B}
                                                        \label{2.11}
\eeq
then the bilinear part of the action $S+S_{gf}$ and the operator $\hat{H}$
has especially simple (minimal) structure
$$
\left(S + S_{gf}\right)^{(2)}
=\int d^4 x \sqrt{-g}\; {\omega} \hat{H} {\omega}^T
$$
\beq
\hat{H}=\hat{K}\Box+\hat{L}_{\rho}\nabla^{\rho}+\hat{M}    \label{2.12}
\eeq
Here $\omega=\left(\bar{h}_{\mu\nu},\;h,\; \varphi\right)$ and $T$ means
transposition. The components of $\hat{H}$ have the form
\[
\hat{K}=\left(
\begin{array}{ccc}
              \frac{B}{4} \delta^{\mu \nu \alpha \beta} & 0 & 0\\
              0 & -\frac{B}{16} & -\frac{B_1}{4} \\
              0 & -\frac{B_1}{4} & \frac{B_1^2}{2B} -A
\end{array}
\right)
\]
\[
\hat{L}^{\lambda}=\left(
\begin{array}{ccc}
              \frac{B_1}{4} \left(\delta^{\mu \nu \alpha \beta}
g^{\tau \lambda} +
2 g^{\nu \beta}\left(g^{\mu \tau } g^{\alpha \lambda }
             - g^{\alpha \tau } g^{\mu \lambda }\right)\right)
         & - \frac{B_1}{4} g^{\mu \tau} g^{\nu \lambda}
       & \left( \frac{B_2}{2}-A \right) g^{\mu \tau} g^{\nu \lambda}\\
           \frac{B_1}{4} g^{\alpha \tau} g^{\beta \lambda}
            & -\frac{B_1}{16} g^{\tau \lambda}
        & \left(\frac{A}{4} -\frac{5}{8} B_2 \right) g^{\tau \lambda}\\
   \left( A - \frac{B_2}{2}\right) g^{\alpha \tau} g^{\beta \lambda}
            & \left( \frac{B_2}{8}-\frac{A}{4}\right) g^{\tau \lambda}
            & \left(\frac{B_1^2}{2B} -A\right)_1 g^{\tau \lambda}
\end{array}
\right) (\nabla_{\tau}\phi)
\]
{\small
\[
\hat{M}=
\left(
\begin{array}{ccc}
              \begin{array}{l}
                   \delta^{\mu \nu \alpha \beta}\left( \frac{B_1}{2}
(\Box \phi)  + \left( \frac{B_2}{2}-\frac{A}{4} \right) (\nabla \phi)^2
                   -\frac{C}{4} \right)
               \\  + g^{\nu \beta}\left(
          -B_1\left( \nabla^{\mu} \nabla^\alpha \phi  \right)
   +\left( A-B_2 \right)(\nabla^\mu \phi)(\nabla^\alpha \phi)\right)
           \\ +\frac{B}{4}\left( -\delta^{\mu \nu \alpha \beta} R
 + 2 g^{\nu \beta} R^{\mu \alpha}+2R^{\mu \alpha \nu \beta} \right)
           \end{array}
           \!\!\!\! &  \!\!\!\! 0
           \!\!\!\! &  \!\!\!\! \begin{array}{l}
           \frac{B_2}{2}\left( \nabla^{\mu} \nabla^{\nu} \phi \right)
              \\
               + \left( \frac{B_3}{2} - \frac{A_1}{2} \right)
                                  (\nabla^\mu \phi)(\nabla^\nu \phi)
              \\
                - \frac{B_1}{2}R^{\mu \nu}
              \end{array}

           \\
          \!\!\!\! & \!\!\!\!
          \!\!\!\! & \!\!\!\!
           \\
\frac{B_1}{4}\left( \nabla^{\alpha} \nabla^{\beta} \phi \right)
              + \frac{B_2}{4} (\nabla^\alpha \phi)(\nabla^\beta \phi)
            & \!\! \frac{C}{16}
            & \!\! \begin{array}{l}

              -\frac{3}{8} B_2 (\Box \phi)
              \\
        + \left( \frac{A_1}{8} - \frac{3}{8}B_3 \right)(\nabla \phi)^2
              \\
              + \frac{B_1}{8} R + \frac{C_1}{4}
              \end{array}
\\
          \!\!\!\! & \!\!\!\!
          \!\!\!\! & \!\!\!\!

           \\

              A\left( \nabla^{\alpha} \nabla^{\beta} \phi \right)
             + \frac{A_1}{2} (\nabla^\alpha \phi)(\nabla^\beta \phi)
              - \frac{B_1}{2}R^{\alpha \beta}
            & \begin{array}{l}
              -\frac{A}{4} (\Box \phi)
            \\
            + \frac{A_1}{8} (\nabla \phi)^2
            \\
            + \frac{B_1}{8} R + \frac{C_1}{4}
              \end{array}
            &
              \begin{array}{l}
              -A_1(\Box \phi)
              \\
              -\frac{A_2}{2}(\nabla \phi)^2
              \\
               + \frac{B_2}{2} R + \frac{C_2}{2}
              \end{array}

\end{array}
\right)
\]
}
The next problem is to separate the divergent part of $\Tr\ln\hat{H}$.
To make this we rewrite this trace in a following way.
\beq
\Tr \ln\hat{H}  =\Tr\ln\hat{K}+
\Tr\ln\left(\hat{1}\Box + \hat{K}^{-1} \hat{L}^{\mu}\nabla_\mu
+\hat{K}^{-1}\hat{M} \right)                            \label{2.15}
\eeq
One can notice that the first term does not give contribution to
the divergences. Let us explore  the second term which has standard
minimal form and can be easily estimated with the use of standard
Schwinger-DeWitt method \cite{DW,hove} (see also \cite{book} for
technical introduction and more complete references).

The bilinear form of the ghost action also has the minimal structure
\beq
\hat{H}_{ghost}=
g^{\mu \alpha}\Box+\gamma(\nabla^{\alpha}\phi)\nabla^{\mu}
+ \gamma (\nabla^{\mu} \nabla^{\alpha} \phi) + R^{\mu \alpha} \label{2.16}
\eeq
and it's contribution to the divergences can be easily derived with the use of
the standard methods.

Summing up both contributions we find that the one-loop divergences
have the form (\ref{2.2}) that is in a full accord with the power counting
consideration. The coefficient functions $c$ have the form
$$
c_w=
     {{43}\over {120}} - {{{{B_{1}}^2}}\over X},\;\;\;\;\;\;\;\;
\;\;\;\;\;\;\;\;\;\;\;\;X=2A\,B\,-3B_{1}^{2}
$$$$
c_r=
     {1 \over {{X^2}}}
\left[
      {1 \over {72}}
         (76\,{A^2}\,{B^2} - 132\,A\,B\,{{B_{1}}^2} + 171\,{{B_{1}}^4})
      - {1 \over 6}
         (B\,( 2\,A\,B + 9\,{{B_{1}}^2} ) \,B_{2})
      + {1 \over 2}
         ({B^2}\,{{B_{2}}^2})
\right]
$$$$
c_4=
     {1 \over {{X^3}}}
\left[
      {1 \over {24\,{B^2}}}
       (128\,{A^4}\,{B^4}
       + 8\,A\,{B^5}\,{{A_{1}}^2}
       - 144\,{A^2}\,{B^4}\,A_{1}\,B_{1}
       - 1552\,{A^3}\,{B^3}\,{{B_{1}}^2}
\right.
$$$$
       + 36\,{B^4}\,{{A_{1}}^2}\,{{B_{1}}^2}
       + 312\,A\,{B^3}\,A_{1}\,{{B_{1}}^3}
       + 5208\,{A^2}\,{B^2}\,{{B_{1}}^4}
       - 6786\,A\,B\,{{B_{1}}^6}
       + 3159\,{{B_{1}}^8})
$$$$
     + {{B_{2}} \over {6\,B}}
       ( -80\,{A^3}\,{B^3}
       - 6\,{B^4}\,{{A_{1}}^2}
       - 24\,A\,{B^3}\,A_{1}\,B_{1}
       + 402\,{A^2}\,{B^2}\,{{B_{1}}^2}
       - 54\,{B^2}\,A_{1}\,{{B_{1}}^3}
       - 810\,A\,B\,{{B_{1}}^4}
$$$$
\left.
         + 459\,{{B_{1}}^6} )
         +  2\,( 2\,{A^2}\,{B^2}
         + 3\,{B^2}\,A_{1}\,B_{1}
         + 12\,A\,B\,{{B_{1}}^2}
         - 9\,{{B_{1}}^4} ) \,{{B_{2}}^2}
         - 6\,A\,{B^2}\,{{B_{2}}^3}
\right]
$$$$
c_5=
     {1 \over {{X^2}}}
\left[
      {1 \over {12\,B}}
        (4\,A\,{B^3}\,A_{1}
        - 128\,{A^2}\,{B^2}\,B_{1}
        + 18\,{B^2}\,A_{1}\,{{B_{1}}^2}
        + 270\,A\,B\,{{B_{1}}^3}
        - 225\,{{B_{1}}^5})
\right.
$$$$
\left.
     + {{B_{2}} \over 2}
        ( -2\,{B^2}\,A_{1}
        + 4\,A\,B\,B_{1}
        + 3\,{{B_{1}}^3} )
\right]
$$$$
c_6=
     {1 \over {{X^2}}}
\left[
      {{B_{1}} \over {{B^2}}}
        ( 8\,A\,{B^3}\,A_{1}
        - 4\,{A^2}\,{B^2}\,B_{1}
        - 6\,{B^2}\,A_{1}\,{{B_{1}}^2}
        + 22\,A\,B\,{{B_{1}}^3}
        -  15\,{{B_{1}}^5} )
\right.
$$$$
\left.
      + {{B_{2}} \over B}
         ( -8\,{A^2}\,{B^2} + 2\,{B^2}\,A_{1}\,B_{1} + 4\,A\,B\,{{B_{1}}^2} -
           3\,{{B_{1}}^4} )
           - 4\,A\,B\,{{B_{2}}^2}
\right]
$$$$
c_7=
     {1 \over {{X^2}}}
\left[
       {2 \over {3\,B}}
         ( 26\,{A^2}\,{B^2}\,{\rm C}
         - 85\,A\,B\,{\rm C}\,{{B_{1}}^2}
         + 63\,{\rm C}\,{{B_{1}}^4}
         + 3\,B\,{\rm C}\,{{B_{1}}^2}\,B_{2}
\right.
$$$$
\left.
          + 8\,A\,{B^2}\,B_{1}\,{\rm C}_{1}
        - 6\,{B^2}\,B_{1}\,B_{2}\,{\rm C}_{1} )
      + {B\,{\rm C}_{2} \over 6}
         ( -2\,A\,B - 9\,{{B_{1}}^2} + 6\,B\,B_{2} )
\right]
$$$$
c_8=
     {1 \over {32\,{B^4}\,{X^4}}}
\left[
      2560\,{A^6}\,{B^6}
      - 448\,{A^3}\,{B^7}\,{{A_{1}}^2}
      + 16\,{B^8}\,{{A_{1}}^4}
      + 256\,{A^4}\,{B^7}\,A_{2}
\right.
$$$$
      + 2432\,{A^4}\,{B^6}\,A_{1}\,B_{1}
      - 22528\,{A^5}\,{B^5}\,{{B_{1}}^2}
      + 1280\,{A^2}\,{B^6}\,{{A_{1}}^2}\,{{B_{1}}^2}
      - 1152\,{A^3}\,{B^6}\,A_{2}\,{{B_{1}}^2}
$$$$
      - 17120\,{A^3}\,{B^5}\,A_{1}\,{{B_{1}}^3}
      + 96\,{B^6}\,{{A_{1}}^3}\,{{B_{1}}^3}
      + 81568\,{A^4}\,{B^4}\,{{B_{1}}^4}
      - 880\,A\,{B^5}\,{{A_{1}}^2}\,{{B_{1}}^4}
$$$$
      + 1856\,{A^2}\,{B^5}\,A_{2}\,{{B_{1}}^4}
      + 42016\,{A^2}\,{B^4}\,A_{1}\,{{B_{1}}^5}
      - 158592\,{A^3}\,{B^3}\,{{B_{1}}^6}
      + 168\,{B^4}\,{{A_{1}}^2}\,{{B_{1}}^6}
$$$$
      - 1248\,A\,{B^4}\,A_{2}\,{{B_{1}}^6}
      - 43512\,A\,{B^3}\,A_{1}\,{{B_{1}}^7}
      + 176976\,{A^2}\,{B^2}\,{{B_{1}}^8}
      + 288\,{B^3}\,A_{2}\,{{B_{1}}^8}
$$$$
      + 16416\,{B^2}\,A_{1}\,{{B_{1}}^9}
      - 108144\,A\,B\,{{B_{1}}^{10}}
      + 28323\,{{B_{1}}^{12}}
      - 2048\,{A^5}\,{B^6}\,B_{2}
$$$$
      + 224\,{A^2}\,{B^7}\,{{A_{1}}^2}\,B_{2}
      + 128\,{A^3}\,{B^7}\,A_{2}\,B_{2}
      + 4480\,{A^3}\,{B^6}\,A_{1}\,B_{1}\,B_{2}
      - 15248\,{A^2}\,{B^5}\,A_{1}\,{{B_{1}}^3}\,B_{2}
$$$$
      + 14208\,{A^4}\,{B^5}\,{{B_{1}}^2}\,B_{2}
      - 1408\,A\,{B^6}\,{{A_{1}}^2}\,{{B_{1}}^2}\,B_{2}
      - 448\,{A^2}\,{B^6}\,A_{2}\,{{B_{1}}^2}\,B_{2}
      - 192\,{B^7}\,{{A_{1}}^3}\,B_{1}\,B_{2}
$$$$
      - 28416\,{A^3}\,{B^4}\,{{B_{1}}^4}\,B_{2}
      + 744\,{B^5}\,{{A_{1}}^2}\,{{B_{1}}^4}\,B_{2}
      + 480\,A\,{B^5}\,A_{2}\,{{B_{1}}^4}\,B_{2}
      +  13680\,A\,{B^4}\,A_{1}\,{{B_{1}}^5}\,B_{2}
$$$$
      + 9696\,{A^2}\,{B^3}\,{{B_{1}}^6}\,B_{2}
      - 144\,{B^4}\,A_{2}\,{{B_{1}}^6}\,B_{2}
      - 2628\,{B^3}\,A_{1}\,{{B_{1}}^7}\,B_{2}
      + 21672\,A\,{B^2}\,{{B_{1}}^8}\,B_{2}
$$$$
      - 15444\,B\,{{B_{1}}^{10}}\,B_{2}
      + 1792\,{A^4}\,{B^6}\,{{B_{2}}^2}
      + 192\,A\,{B^7}\,{{A_{1}}^2}\,{{B_{2}}^2}
      + 60624\,{A^2}\,{B^4}\,{{B_{1}}^4}\,{{B_{2}}^2}
$$$$
      - 19328\,{A^3}\,{B^5}\,{{B_{1}}^2}\,{{B_{2}}^2}
      + 576\,{B^6}\,{{A_{1}}^2}\,{{B_{1}}^2}\,{{B_{2}}^2}
      + 7776\,A\,{B^5}\,A_{1}\,{{B_{1}}^3}\,{{B_{2}}^2}
      - 928\,{A^2}\,{B^6}\,A_{1}\,B_{1}\,{{B_{2}}^2}
$$$$
      - 6984\,{B^4}\,A_{1}\,{{B_{1}}^5}\,{{B_{2}}^2}
      - 69120\,A\,{B^3}\,{{B_{1}}^6}\,{{B_{2}}^2} +
      25380\,{B^2}\,{{B_{1}}^8}\,{{B_{2}}^2}
      + 128\,{A^3}\,{B^6}\,{{B_{2}}^3}
$$$$
\left.
      - 1152\,A\,{B^6}\,A_{1}\,B_{1}\,{{B_{2}}^3}
      - 3072\,{A^2}\,{B^5}\,{{B_{1}}^2}\,{{B_{2}}^3}
      + 3888\,{B^3}\,{{B_{1}}^6}\,{{B_{2}}^3}
      + 576\,{A^2}\,{B^6}\,{{B_{2}}^4}
\right]
$$$$
     + {{B_{3}} \over {4\,{B^2}\,{X^2}}}
\left[
         -10\,A\,{B^3}\,A_{1}
        - 24\,{A^2}\,{B^2}\,B_{1}
        + 19\,{B^2}\,A_{1}\,{{B_{1}}^2}
        +  36\,A\,B\,{{B_{1}}^3}
\right.
$$$$
\left.
        - 6\,{{B_{1}}^5}
        + 8\,A\,{B^2}\,B_{1}\,B_{2}
        - 36\,B\,{{B_{1}}^3}\,B_{2}
\right]
$$$$
c_9=
     {1 \over {8\,{B^3}\,{X^3}}}
\left[
       -16\,{A^3}\,{B^5}\,A_{1}
       + 8\,{B^6}\,{{A_{1}}^3}
       + 576\,{A^4}\,{B^4}\,B_{1}
       - 16\,A\,{B^5}\,{{A_{1}}^2}\,B_{1}
\right.
$$$$
      + 160\,{A^2}\,{B^4}\,A_{1}\,{{B_{1}}^2}
      - 2960\,{A^3}\,{B^3}\,{{B_{1}}^3}
      + 12\,{B^4}\,{{A_{1}}^2}\,{{B_{1}}^3}
      - 324\,A\,{B^3}\,A_{1}\,{{B_{1}}^4}
$$$$
      + 5400\,{A^2}\,{B^2}\,{{B_{1}}^5}
      + 90\,{B^2}\,A_{1}\,{{B_{1}}^6} .
      - 4176\,A\,B\,{{B_{1}}^7}
      + 1107\,{{B_{1}}^9}
$$$$
      + 56\,{A^2}\,{B^5}\,A_{1}\,B_{2}
      + 960\,{A^3}\,{B^4}\,B_{1}\,B_{2}
      - 48\,{B^5}\,{{A_{1}}^2}\,B_{1}\,B_{2}
      - 144\,A\,{B^4}\,A_{1}\,{{B_{1}}^2}\,B_{2}
$$$$
      - 4032\,{A^2}\,{B^3}\,{{B_{1}}^3}\,B_{2}
      + 234\,{B^3}\,A_{1}\,{{B_{1}}^4}\,B_{2}
      + 6228\,A\,{B^2}\,{{B_{1}}^5}\,B_{2}
      - 3186\,B\,{{B_{1}}^7}\,B_{2}
$$$$
\left.
      + 48\,A\,{B^5}\,A_{1}\,{{B_{2}}^2}
      - 240\,{A^2}\,{B^4}\,B_{1}\,{{B_{2}}^2}
      + 216\,A\,{B^3}\,{{B_{1}}^3}\,{{B_{2}}^2}
      - 108\,{B^2}\,{{B_{1}}^5}\,{{B_{2}}^2}
\right]
     - {{B_{3}}\over B}
$$$$
c_{10}=
     {1 \over {8{B^2}{X^2}}}
\left[
      4B^3A_1(B A_1
      - 4AB_1)
      + 12 B^2B_1^2 (18A^2
     - A_1B_1)+B_1^4(387 B_1^2
     - 528AB)
\right]
    - {{B_{2}}\over B}
$$$$
c_{11}=
     {1 \over {4\,{B^3}\,{X^3}}}
\left[
     (64\,{A^4}\,{B^4}
     + 16\,{A^2}\,{B^4}\,A_{1}\,B_{1}
     - 832\,{A^3}\,{B^3}\,{{B_{1}}^2}
    - 40\,{B^4}\,{{A_{1}}^2}\,{{B_{1}}^2}
\right.
$$$$
     + 16\,A\,{B^4}\,A_{2}\,{{B_{1}}^2}
     + 40\,A\,{B^3}\,A_{1}\,{{B_{1}}^3}
     + 3176\,{A^2}\,{B^2}\,{{B_{1}}^4}
     - 24\,{B^3}\,A_{2}\,{{B_{1}}^4}
     - 24\,{B^2}\,A_{1}\,{{B_{1}}^5}
$$$$
     - 4356\,A\,B\,{{B_{1}}^6}
      + 2070\,{{B_{1}}^8}
      - 96\,{A^3}\,{B^4}\,B_{2}
      + 40\,A\,{B^4}\,A_{1}\,B_{1}\,B_{2}
      + 160\,{A^2}\,{B^3}\,{{B_{1}}^2}\,B_{2}
$$$$
      +  84\,{B^3}\,A_{1}\,{{B_{1}}^3}\,B_{2}
      - 420\,A\,{B^2}\,{{B_{1}}^4}\,B_{2}
      + 234\,B\,{{B_{1}}^6}\,B_{2}
      - 72\,A\,{B^3}\,{{B_{1}}^2}\,{{B_{2}}^2}
      + 36\,{B^2}\,{{B_{1}}^4}\,{{B_{2}}^2}
   )\,{\rm C}
$$$$
  + (  - 16\,{A^2}\,{B^5}\,A_{1}
      + 96\,{A^3}\,{B^4}\,B_{1}
      + 80\,{B^5}\,{{A_{1}}^2}\,B_{1}
      - 32\,A\,{B^5}\,A_{2}\,B_{1}
      - 56\,A\,{B^4}\,A_{1}\,{{B_{1}}^2}
$$$$
      - 808\,{A^2}\,{B^3}\,{{B_{1}}^3}
      + 48\,{B^4}\,A_{2}\,{{B_{1}}^3}
      - 24\,{B^3}\,A_{1}\,{{B_{1}}^4}
      + 1140\,A\,{B^2}\,{{B_{1}}^5}
$$$$
      - 612\,B\,{{B_{1}}^7}
      - 40\,A\,{B^5}\,A_{1}\,B_{2}
      + 224\,{A^2}\,{B^4}\,B_{1}\,B_{2}
      - 228\,{B^4}\,A_{1}\,{{B_{1}}^2}\,B_{2}
$$$$
      + 108\,A\,{B^3}\,{{B_{1}}^3}\,B_{2}
      + 54\,{B^2}\,{{B_{1}}^5}\,B_{2}
      + 120\,A\,{B^4}\,B_{1}\,{{B_{2}}^2}
      - 36\,{B^3}\,{{B_{1}}^3}\,{{B_{2}}^2}
 )\,{\rm C}_{1}
$$$$
      + (  64\,{A^3}\,{B^5}
      - 20\,{B^6}\,{{A_{1}}^2}
      + 8\,A\,{B^6}\,A_{2}
      - 16\,A\,{B^5}\,A_{1}\,B_{1}
      - 112\,{A^2}\,{B^4}\,{{B_{1}}^2}
$$$$
      - 12\,{B^5}\,A_{2}\,{{B_{1}}^2}
      + 60\,{B^4}\,A_{1}\,{{B_{1}}^3}
      + 120\,A\,{B^3}\,{{B_{1}}^4}
      - 45\,{B^2}\,{{B_{1}}^6}
      - 16\,{A^2}\,{B^5}\,B_{2}
$$$$
\left.
      + 72\,{B^5}\,A_{1}\,B_{1}\,B_{2}
      - 48\,A\,{B^4}\,{{B_{1}}^2}\,B_{2}
      - 72\,{B^3}\,{{B_{1}}^4}\,B_{2}
      - 24\,A\,{B^5}\,{{B_{2}}^2}
  )\,{\rm C}_{2}
\right]
$$$$
   + { {{\rm C}_{3}} \over {2\,{X^2}}}
\left[
     2\,{B^2}\,A_{1}
    - 4\,A\,B\,B_{1}
    - 3\,{{B_{1}}^3}
\right]
$$$$
c_{12}=
     {1 \over {{X^2}}}
\left[
       {1 \over {{B^2}}}
         (20\,{A^2}\,{B^2}\,{{{\rm C}}^2}
         - 56\,A\,B\,{{{\rm C}}^2}\,{{B_{1}}^2}
         + 41\,{{{\rm C}}^2}\,{{B_{1}}^4}
         - 8\,A\,{B^2}\,{\rm C}\,B_{1}\,{\rm C}_{1}
     + { {{B^2}\,{{{\rm C}_{2}}^2}} \over 2}
\right.
$$
\beq
\left.
         + 4\,A\,{B^3}\,{{{\rm C}_{1}}^2}
         + 2\,{B^2}\,{{B_{1}}^2}\,{{{\rm C}_{1}}^2})
        + 2\,B_{1}\,( {\rm C}\,B_{1}
     - 2\,B\,{\rm C}_{1} ) \,{\rm C}_{2}
         + 4\,B\,{\rm C}\,{{B_{1}}^3}\,{\rm C}_{1}
\right]                                          \label{2.17}
\eeq

Let us now make some comments concerning the above result. The one loop
divergences (\ref{2.2}), (\ref{2.17}) essentially depend on the choice of the
functions $A(\phi), B(\phi), C(\phi)$ in the starting action (\ref{0.1}).
In particular, this dependence concerns the $c_w$ and $c_r$ functions,
which correspond to the terms with the second powers in curvature tensor.
The above expressions are valid only in the case
$\;X=2A\,B\,-3B_{1}^{2}\neq 0$.
For $X=0$ the calculational scheme must be modified because of extra
conformal symmetry. In this case
one has to introduce the additional gauge fixing condition for conformal
symmetry.
It is easy to see that if such condition is taken in the form $h=0$ then
the degeneracy of $\hat{K}$ is removed.

The curvature squared terms in  (\ref{2.17}) are in a good accord with
the same terms calculated earlier in \cite{hove}. The direct comparison of
(\ref{2.17}) with the results of other authors is difficult since they have
used different choice of quantum variables.

\section{One-loop finiteness and renormalizability on shell}

And so we observe that the one-loop divergences in the theory under
discussion have rather complicated form, and include all possible structures
of the action (\ref{2.2}). In this respect the theory is similar to
General Relativity, where all possible counterterms also appear
\cite{hove}. It is well
 known that in the last case all the counterterms disappear on mass shell.
Therefore it is interesting to consider the divergences (\ref{2.17}) when
the equations of motion (\ref{2.1}) are taken into account.
Then, with the use of (\ref{2.5}) and (\ref{2.6}) we obtain the one
loop divergences in the form
\beq
\Gamma_{on-shell, div}^{1-loop}= - \frac{1}{\epsilon}
\int d^4x \left[ k_w C^2 + k_{rr} R^2 + k_r R + k_l \right]   \label{3.1}
\eeq
where $\epsilon = (4\pi)^2 \;(n-4)$ and the values of coefficients are
$$
k_l(\phi) =
{\it c_{12}} + {\it c_{10}}\,{w^2} + {\it c_{11}}\,y + {\it c_9}\,w\,y
+ {\it c_8}\,{y^2}
$$$$
k_r(\phi)=
{\it c_7} - f\,u + {\it c_5}\,w + {{f\,w}\over 2} + {\it c_{11}}\,x +
{\it c_9}\,w\,x +
$$$$
+  {\it c_4}\,y + {{{\it c_6}\,y}\over 2} - {{f\,r\,y}\over 2} + 2\,
{\it c_8}\,x\,y +
  2\,{\it c_{10}}\,w\,z + {\it c_9}\,y\,z
$$$$
k_{rr}(\phi) =
{\it c}_r - {{f\,s}\over 3} - f\,t + {\it c_4}\,x + {{{\it c_6}\,x}\over 2} -
  {{f\,r\,x}\over 2} + {\it c_8}\,{x^2} + {\it c_5}\,z + {{f\,z}\over 2} +
  {\it c_9}\,x\,z + {\it c_{10}}\,{z^2}
$$
\beq
k_w(\phi) =
{\it c}_w - {{f\,s}\over 2}                           \label{3.2}
\eeq
and $x,y,z,w,t,u,f$ have been defined in (\ref{2.5}), (\ref{2.6}).
It is very important that the values of $k_{w,rr,r,l}$ do not depend
on the choice of gauge fixing. From this follows that the finite solutions
which we find below are independent on gauge and therefore good defined.

The divergences in the higher derivative
dimension-less sector do not depend on the
dimensional function $C(\phi)$ and therefore can be analyzed independently
on the others.
Let us start the search of finite solutions with the higher derivative
structures and consider the equations $k_w=o$ and $k_{rr}=0$.
In order to find the solutions we must solve these differential equations
that looks extremely difficult problem. However one can see that the
 divergent coefficients (\ref{2.17}) possess some homogeneouty.
Taking this into account one can successfully find the finite
solutions. In fact we have found three finite solutions of power type
\beq
A(\phi)=a\phi^m,\; \;\;\;\;\;\;\;\;\;\;\;\;\;\;\;
B(\phi)=b\phi^{m+2}                                   \label{3.3}
\eeq
with different real values of $m$ and of the ratio $\frac{a}{b}$.
\beq
   (m_1,\frac{a_1}{b_1})=
(-23.4851... ,1413.45...),\;\;\;\;\;\;\;\;\;\;\;\;
\; \;\;\;\;                                          \label{3.41}
\eeq
\beq
(m_2,\frac{a_2}{b_2})= (-0.526300...,2.72924...),
\;\;\;\;\;\;\;\;\;\;\;\;\; \;\;\;\;                    \label{3.42}
\eeq
\beq
(m_3,\frac{a_3}{b_3})=
(-0.317820...,4.09345...)                             \label{3.43}
\eeq
The conditions $k_w=o$ and $k_{rr}=0$ do not fix the values of the
constants $a, b$ but only their ratio. Let us notice that the algebraic
equation for $m$ is of fifth order and therefore the fact of existence of
real solution is independent of the numerical details in the expression
(\ref{2.17}).
Thus we have found the form of the functions $A(\phi)$ and $B(\phi)$
for which our dilaton model without cosmological term $C(\phi)$
is finite on shell. Note that $C(\phi)$ term doesn't give contributions to
the higher derivative counterterms due to it's dimension. However,
if this term is included, situation becomes a little bit more
 complicated.

 If we substitute the  expressions (\ref{3.3})
into the classical action (\ref{0.1}), then the
 term, linear in curvature, is equal to zero.
Thus the theory under consideration
can be renormalizable on shell only if $k_r=0$. Therefore we must choose
$C(\phi)$ in such a way that the counterterm, linear in curvature, is lacking
on shell.

If we choose $C(\phi)=c \phi^{2(m+2)}$ (that corresponds to the same power
of $\phi$ in both $C(\phi)$ and counterterm) then
the classical action of the theory has the form
\beq
S= \int d^4 x \sqrt{-g} \left[ - c \phi^{2(m+2)} \right]   \label{3.5}
\eeq
and the on shell divergences are
\beq
\Gamma_{on-shell, div}^{1-loop}=-\frac{1}{16 \pi^2 (n-4)}
\int d^4x \left[
     s_1\frac{c}{b} \phi^{m+2} R + s_2 \frac{c^2}{b^2}\phi^{2(m+2)} \right]
                                                          \label{3.6}
\eeq
where $s_1, s_2$ have the approximate numerical values
(here and below we omit dots)
$$
(s_1, s_2)=(1.45386, 0.05772),\;\;\;
(-0.55182, -3.95364),\;\;\;
(-0.105262, -3.06052)
$$
for the solutions (\ref{3.41})-(\ref{3.43}) correspondingly.
Thus for this choice of $C(\phi)$
the theory is not renormalizable even on mass shell.
Hence we have to look for another values of the power.

Substituting (\ref{3.3}) into the expression for $k_r$ we obtain
the ordinary differential equation for $C(\phi)$.
\beq
\lambda_0\phi^3C_3+\lambda_1\phi^2C_2+\lambda_2\phi C_1+
\lambda_3 C=0                                              \label{3.7}
\eeq
Here the indices stand for the derivatives with respect to $\phi$ (\ref{1.6}).
For three versions of
$(m,\frac{a}{b})$ given in (\ref{3.41}), (\ref{3.42}), (\ref{3.43})
correspondingly, the constants $\lambda_{0,1,2,3}$ have the following
values:
\beq
\lambda_0=-2.49128 ,\;\;\;\;\lambda_1=  -813.35,\;\;\;\;
\lambda_2= -6170.03 ,\;\;\;\;\lambda_3= 1310940            \label{3.8}
\eeq
\beq
\lambda_0= 1.25083,\;\;\;\;\lambda_1=-7.41961 ,\;\;\;\;
\lambda_2=13.6904 ,\;\;\;\;\lambda_3=-4.71357    \label{3.9}
\eeq
\beq
\lambda_0= 2.94653,\;\;\;\;\lambda_1=-23.1234 ,\;\;\;\;
\lambda_2=60.081 ,\;\;\;\;\lambda_3=-50.1804      \label{3.10}
\eeq

The equations (\ref{3.7}) can be easily solved in the form
\beq
C_i(\phi)=L_{i1}\phi^{k_{i1}}+L_{i2}\phi^{k_{i2}}+L_{i3}\phi^{k_{i3}}
                                                            \label{3.11}
\eeq
where $C_i(\phi)$ corresponds to $m_i$  and
 $L_{11,12,...,33}$ are arbitrary integration constants and
the following powers  correspond to coefficients (\ref{3.8}),
(\ref{3.9}), (\ref{3.10})
\beq
k_{11}=35.4101,\;\;\;\;\;\;\;\;
k_{12}=-47.7636,\;\;\;\;\;\;\;\;
k_{13}=-311.125                                                \label{3.111}
\eeq
\beq
k_{21}=5.77717,\;\;\;\;\;\;\;\;
k_{22}=0.222461,\;\;\;\;\;\;\;\;
k_{23}=2.93212                                                 \label{3.121}
\eeq
\beq
k_{31}=3.36417,\;\;\;\;\;\;\;\;
k_{32}=0.752027,\;\;\;\;\;\;\;\;
k_{33}=6.73149                                                 \label{3.131}
\eeq

The above solutions give the form of the function $C(\phi)$ which provide
the absence of the non-renormalizable $k_r R $ type counterterm.
The only structures which survive on shell in both classical action and
counterterms are the potential ones. In this respect the theory under
consideration is sharing the corresponding property of the Einstein gravity,
where the only cosmological term remains when one uses the equations of motion.
In the Einstein gravity with cosmological term
this leads to the renormalizability of the
theory on shell and enables one to consider the renormalization group
equation for the cosmological constant.

Let us  do the same and
consider the renormalization of the potential function $C(\phi)$.
We consider only the simple
case of $C(\phi)=L\phi^{k}$ that is take two of
$L$'s equal to zero, and the third arbitrary. The renormalization
of the potential function follows from (\ref{3.3}), (\ref{3.2}), (\ref{2.17})
and has the form
\beq
L^{(0)}\;(\phi^{(0)})^{k} = \mu^{n-4} \left[ L \;\phi^{k}+
\frac{1}{\varepsilon}\;\frac{Q\;L^2}{b^2}\;\phi^{2k-2m-4} \right]  \label{3.12}
\eeq

Here we have included the factors of dimensional parameter $\mu$ related with
the use of dimensional regularization.
$Q$ is some number, which depends on the ratio of $a$ and $b$. The values of
$Q$ which correspond to the $k$'s from (\ref{3.111}) - (\ref{3.131})
and $m$'s from (\ref{3.41}) are
\beq
Q_{11}=10.2275,\;\;\;\;\;\;\;\;
Q_{12}=-1.51342,\;\;\;\;\;\;\;\;
Q_{13}=1249.04                                                \label{3.211}
\eeq
\beq
Q_{21}=19.7562,\;\;\;\;\;\;\;\;
Q_{22}=26.9177,\;\;\;\;\;\;\;\;
Q_{23}=-1.39468                                                 \label{3.221}
\eeq
\beq
Q_{31}=-1.42338,\;\;\;\;\;\;\;\;
Q_{32}=268.66,\;\;\;\;\;\;\;\;
Q_{33}=664.993                                                 \label{3.231}
\eeq

 It is easy to see that the potential type divergences
can not be removed by the transformation of $L$ but
only by the following renormalization transformation of
scalar field $\phi$.
\beq
\phi^{(0)}=\mu^{\frac{n-4}{k}} \left[ \phi+\frac{1}{\varepsilon}\;
\frac{Q(a/b)\;L}{k\;b^2} \;\phi^{k-2m-3} \right]               \label{3.13}
\eeq

The relation (\ref{3.12}) does not enable us to find the dimensions of both
$L$ and $\phi$. Since all three terms in (\ref{3.12}) have (if $n=4$)
dimension 4, we can easily obtain  two equations for dimensions $d_L$, $d_b$
(dimension of constant $b$) and $d_\phi$.
$$
d_L + k\;d_\phi = 4
$$
\beq
d_L + (k-m-2)\; d_\phi - d_b = 2                          \label{3.14}
\eeq
Another equation for these dimensions comes from the starting
action
\beq
d_b + (m+2)\; d_\phi  = 2                          \label{3.103}
\eeq
together with $d_a = d_b$. However the last three equations are dependent
and thus we only can express the dimensions of $a, b, L$ via $d_\phi$.

The renormalization relation (\ref{3.13}) together with (\ref{3.14}) and
(\ref{3.103}) enables us to explore the  renormalization group equations for
the effective charges. he renormalization group function for the scalar field
$\phi$ is defined in a standard way and has the form
\beq
\gamma_\phi = \mu\;\frac{d\;\phi}{d\mu} =
- \frac{n-4}{k}\;\phi +
\frac{1}{(4\pi)^2}\;
\frac{Q(a/b)\;L}{b^2\;k^2} \;(k-2m-4)\;\phi^{k-2m-3}
\label{3.14x}
\eeq
Since we consider the four dimensional theory the first term in the $rhs$ can
be omitted
and thus we arise at the following  renormalization group equations for
$L(t)$ and $\phi(t)$
$$
(4\pi)^2\;\frac{d L}{dt} = (k\;d_\phi  - 4)\;L
$$
\beq
(4\pi)^2\;\frac{d\phi }{dt} = \frac{Q(a/b)\;L}{b^2\;k^2}
\;(k-2m-4)\;\phi^{k-2m-3} - d_\phi \;\phi                 \label{3.14a}
\eeq
Indeed one can easily write the similar equations for $a(t)$ and $b(t)$.
Since these constants are not renormalized, their values depend on scale
only due the classical dimensions. Since all these dimensions
are ill defined the
equations looks rather artificial. The only way to extract  some information
is to consider the quantities with definite dimension.

One can see that in the theory under consideration all the on shell
divergences can be removed by the renormalization of the dilaton field.
As a result there are the nontrivial renormalization group equations
for $\phi$ which acquire the anomalous dimension.
The scale dependence of the parameters $a, b, L$  and the field
$\phi$ leads to the effective running of the cosmological constant.
To see this let us consider the effective potential of the scalar field.
Upon the normalization conditions are introduced, the effective potential
has the form
\beq
V_{eff} = \sum \left\{ b\;\phi^{m+2} \;R\; \left[ 1 + (m+2)\;\gamma\; \ln
\frac{M_i(\phi)}{\mu}
\right] + L\; \phi^k \left[ 1 + k\;\gamma\; \ln \frac{M_i(\phi)}{\mu}
\right] \right\}                                    \label{3.115}
\eeq
where $M_i(\phi)$ are the effective masses, which are the eigenvalues of the
operators (\ref{2.12}) and  (\ref{2.16}) and the algebraic summation
is performed according to rule (\ref{2.9}). $\mu$ is the
dimensional parameter of renormalization, and $\gamma$ is the
renormalization group function for dilaton (\ref{3.14x}).
Since we are interested in
 the qualitative scaling behavior of the Newtonian and cosmological constants
the logarithmic corrections in (\ref{3.115}) are not relevant and one can deal
with the renormalization group improved classical potential of the theory.
\beq
V_{imp}\; = \;  b(t)\;\phi^{m+2}(t)\;R\;+\;L(t)\; \phi^k(t)   \label{3.116}
\eeq
One can easily see that for some of the solutions (\ref{3.111}) -
(\ref{3.131}), (\ref{3.41}) - (\ref{3.43}) for the potential (\ref{3.116})
it is satisfied the criteria of the second order phase transition
\beq
\frac{d V_{imp}}{d \phi} = 0, \; \; \; \; \; \; \; \; \;
 \frac{d^2 V_{imp}}{d \phi^2} > 0                      \label{3.117}
\eeq
The first of (\ref{3.117}) leads to the relation between critical values
of scalar field and curvature
\beq
R_c = - \frac{Lk}{b(m+2)}\; \phi_c^{k-m-2}           \label{3.118}
\eeq
and the second impose the restriction $k - m > 2$. The last condition is
satisfied for the models with $k_{11}, k_{21}, k_{23}, k_{31}, k_{33}$
(\ref{3.111}) - (\ref{3.131}) and corresponding values of $m$ in
(\ref{3.41}) - (\ref{3.43}). Note that the stability of vacuum at classical
level require positive $k, m+2, b$ and negative $L$. Thus the value
of $k_{11}$ is not compatible with physical requirements. However another four
models are and thus our theory allow the second order phase transitions.
In the point of minima $\phi_0$ the renormalization group improved
classical potential
has the form of the Hilbert - Einstein action
\beq
S_{min}=\int d^4x\sqrt{-g}\;\{-\frac{1}{G_{ind}}\;R+\Lambda_{ind}\}
                                                          \label{3.119}
\eeq
where the induced values of Newtonian $G_{ind}$ and cosmological
$\Lambda_{ind}$ constants are defined as
$G_{ind}^{-1}=b\;\phi_0^{m+2}$ and $\Lambda_{ind} = L\;\phi_0^k$.
Since the parameters $b$ and $L$ are not renormalized the scaling dependence
of $G_{ind}$ and $\Lambda_{ind}$ is caused by the only renormalization of
dilaton (\ref{3.13}) and by the classical dimension of the constants,
which are well defined in the case. Since the classical dimension will
surely dominate in the renormalization group equations for
$G_{ind}$ and $\Lambda_{ind}$, it is reasonable to rewrite the renormalization
 group equation  (\ref{3.14a}) in terms of dimension-less parameter
$\eta(t) = \Lambda_{ind}(t)\;G_{ind}^2(t)$. The equation for $\eta(t)$
has remarkably simple form
\beq
(4\pi)^2\;\frac{d \eta }{dt} =
\frac{(k-2m-4)^2}{k^2}\;Q(a/b) \;\eta^2                     \label{3.120}
\eeq
which indicate to the standard asymptotical behavior of this parameter.
And so the scaling behaviour of the induced cosmological constant is
defined by the sign of the number $Q(a/b)$.
One can see from (\ref{3.211}) - (3.231) that for the physically relevant
solutions with $k_{21}, k_{33}$ from (\ref{3.121}), (\ref{3.131})
the value $Q$ is positive. In these models the effective cosmological
constant decrease at low energies.
Thus
we observe that the quantum effects in these versions of
 theory under consideration
describe the vanishing of the cosmological constant in far IR.
As a result
the observable low energy value of the cosmological constant
is small as compared with the high energy value.
On the countrary, for
$ k_{23}, k_{31}$  (\ref{3.121}), (\ref{3.131}) the sign of $Q$
is negative and the effective cosmological
constant decrease at high energy scale.

And so we have found that the model (\ref{0.1}) leads to the effective running
of the cosmological constant if the last is measured in the units of the
Newtonian constants. Earlier the running of these couplings has been in the
framework of Higher derivative models \cite{merc}, in the gauge models on
external classical background \cite{lam} and also in the higher derivative
dilaton model \cite{ejos}. The interesting property of our model is that
here we observe qualitatively different asymptotical regims. In this respect
the theory of one loop quantum dilaton gravity share the features of the higher
derivative dilaton model \cite{ejos}.

\section{Interaction with matter fields}

In the previous sections we have shown there are some
special versions of  the general dilaton model
(\ref{0.1}) which are  renormalizable on shell at one - loop level.
In spite of that
the expression for counterterms (\ref{2.17}) is very cumbersome (one can truly
say terrible) all those counterterms vanish on shell if the starting functions
are chosen in a special way. Thus the special cases (\ref{2.17}) -
(\ref{2.17}), (\ref{2.17}) - (\ref{3.131}) are sharing the same property of
quantum Einstein gravity (with cosmological constant),
which is also renormalizable on shell \cite{hove,cosm}. However in the last
case the interaction with matter fields leads to the violation of one
loop on shell renormalizability \cite{dene}. In this section we show that for
our dilaton model it is also the case. If the matter fields are included
the arbitrary functions $A(\phi), B(\phi)$ can not be fine tuned in such a way
that the on shell higher derivative counterterms vanish.

Below we consider this in some details. One can suppose for simplicity that
the matter fields action is composed by the vector
fields only, and that the action of matter fields has the standard form and
matter fields do not interact with the dilaton field $\phi$ directly, but only
via metric. Thus we consider the most simple case which corresponds to the
first work in  \cite{dene} when the gravity was described by
the Hilbert - Einstein action. According to \cite{dene}
he renormalizability of the  Einstein - Maxwell system is violated by the
counterterms like $T_{\mu\nu}\;T^{\mu\nu}$ and $R_{\mu\nu}\;T^{\mu\nu}$.
Here $T_{\mu\nu}$ is the Energy - Momentum Tensor of the matter
$$
T_{\mu\nu} = - \frac{2}{\sqrt{-g}}\;\frac{\delta S_m}{\delta g^{\mu\nu}}
$$
which is traceless in four dimensions
$T_{\mu\nu}\;g^{\mu\nu}=0$ that reflects the conformal invariance
property of the matter fields action.

The contributions of matter fields and the ones of "mixed sector"
 to the one loop
counterterms in our metric - dilaton gravity
(after the renormalization in a matter fields sector,
however for vectors it is not necessary \cite{dene}) lead to the following
change of
the counterterms.
 The general form of the divergences (\ref{2.2}) is changed because of
$T_{\mu\nu}\;T^{\mu\nu}$ and $R_{\mu\nu}\;T^{\mu\nu}$ terms,
and moreover  coefficient $c_w$  acquire the addition
\beq
c_w\rightarrow c_w+\frac{N_v}{10} + \frac{N_f}{20}       \label{3.15}
\eeq
The contributions to $c_r$ are forbidden by conformal invariance (see, for
example, \cite{book}) and others are lacking because the matter fields
decouple from dilaton.

Let us now discuss the on shell renormalization which is a little bit more
complicated. The first of the classical equations of motion (\ref{2.1})
acquire the additional term $\frac{1}{2}\;T_{\mu\nu}$ on the $rhs$. However
the second equation (\ref{2.1}) remains unchanged as well as the trace of
the first one. The transfer on shall is performed with the use of the
formulas (\ref{2.4}) - (\ref{2.7}). Since (\ref{2.4}) and (\ref{2.5})
are based just on the second equation (\ref{2.1}) and on the
 the trace of the first one, we conclude that for the theory with matter
(\ref{2.4}) and (\ref{2.5}) are the same as in a pure metric - dilaton gravity.
The detailed analysis show that the equation (\ref{2.7}) is also the same.
However, the expression (\ref{2.6}) changes according to
\beq
sR_{\mu\nu}R^{\mu\nu}\rightarrow sR_{\mu\nu}\;R^{\mu\nu}
-\frac{1}{2 B_1(\phi)}  \;R^{\mu\nu}\;T_{\mu\nu}             \label{3.16}
\eeq
Therefore  there is following contribution of the matter
fields sector to the on shell divergences
\beq
\delta\Gamma = - \frac{1}{2\varepsilon}\int d^4 x \sqrt{-g}\;
\frac{f(\phi)}{B_1(\phi)}\;R^{\mu\nu}\;T_{\mu\nu}            \label{3.17}
\eeq
which has to be added to (\ref{3.1}) together with original
$T_{\mu\nu}\;T^{\mu\nu}$ and $R_{\mu\nu}\;T^{\mu\nu}$ terms.
Moreover one must take into account the numerical change
in $k_w$ which corresponds to (\ref{3.15}) and (\ref{3.2}).

Thus if we try to cancel the higher derivative
on shell divergences by choosing the
functions $A$ and $ B$ we face the more difficult problem than we have met
in Section 5. As it was already pointed out there, the equations $k_w = k_{rr}
= 0$ have real solutions and therefore the $R^2$ type counterterms can
be removed. However the $T_{\mu\nu}\;T^{\mu\nu}$ and $R_{\mu\nu}\;T^{\mu\nu}$
structures survive and we already do not have free parameters to cancel them.
Thus one can observe that the on shell renormalizability is lacking in
our metric - dilaton theory just as in purely metric gravity \cite{dene}.
Moreover the renormalizability is violated by exactly the same two structures
which are related with the traceless  Energy - Momentum Tensor of the matter
fields.

\section{Two special cases}

In this section we briefly discuss two special cases of the theory (\ref{0.1})
which are of special physical interest.
One can  consider this part of the paper as some
kind of quantum gravity phenomenology.
\vskip 2mm

i) Let us consider the theory (\ref{1.12}) that is classically equivalent
to the special version of higher derivative quantum gravity (\ref{1.11}).
The transfer to quantum theory can be  performed by introducing the
generating functional of Green functions. If one introduce the external
source for the auxiliary scalar field $\phi$, then the direct link between
two models (\ref{1.11}) and (\ref{1.12}) will be lost. Therefore if we like
to have such a link, the external source must be introduced for the
metric only. If we consider the effective action in a background gauge,
then the lack of  external source for scalar corresponds to the lack of
the  background scalar field. In this case the only counterterms in the theory
are the ones of the
$c_w, c_r, c_7, c_{12}$ type in (\ref{2.2}) and only one of them,
namely  $c_w$, violate the renormalizability. If we restrict ourselves
by only the conformally flat background, then the theory is renormalizable
and we can construct the renormalization group equations for the
effective couplings  $\alpha(t), G(t), \Lambda(t)$. The study of these
equations show that the theory possesses cosmologically acceptable regime
and, in particular, the quantum effects lead to exponential decrease of
$\alpha(t)$ and $ \Lambda(t)$ at high energies \cite{OV}. Indeed the additional
restriction on the background metric is not completely consistent from formal
point of view. On this way we remove the divergent diagrams with massive
spin two particles by hands. On the other side, in higher
derivative gravity the existence of these particles (which have the
wrong sign of the kinetic term and thus are unphysical) lead to the
well known unitarity problem, so their removal here is not much worst then
the existence. The detailed analysis of the renormalization group equations
in the model (\ref{1.11}) and their cosmological consequences will be given
in \cite{OV}.
\vskip 2mm

ii) Another interesting particular case is related with another extra
condition on the background. One can suppose that the  background scalar field
is varying slowly as compared with the metric, and remove all the terms with
the derivatives of scalar. Then we find that the only types of the
counterterms which survive are the same as in the previous case. The
next natural step is to look for the solutions of the equations $c_w = c_r = 0$
and so construct the theory with renormalizable potential. We have explored
the (\ref{3.3}) form of the functions $A(\phi)$ and $B(\phi)$. It turned out
that the equations for $a, b, m$ following from $c_w = c_r = 0$ condition,
do not have real solutions, and hence this idea doesn't work. At the same
time the above equations for $A(\phi)$ and $B(\phi)$ are ill defined because
the conditions $c_w = c_r = 0$ depend on the choice of gauge fixing
parameters. The possible way to remove this dependence is related with the
use of the Vilkovisky's unique effective action \cite{vilk} which coincide
with the conventional effective action on shell, but differs off shell.
Within this scheme we shall get the equations for $A(\phi)$ and $B(\phi)$
which have the structure similar to above but with different numerical
coefficients. One can suppose that taking into account the  Vilkovisky
corrections to the divergences, we can get the real solutions and to
construct the dilaton theory with renormalizable potential.

\section{Discussion}

In this paper we have considered the different aspects of the one-loop
renormalization in the theory (\ref{0.1}). We have shown that the models
of this type can be divided into two classes - models of one are conformally
equivalent to the general relativity and also to the conformally coupled
to gravity scalar field. The models of second class are conformally
equivalent to the model (\ref{1.1}) with non-constant $\Phi$, and any
model of this type can be related to another one  by some change of variables
together with some change of potential function.

The one-loop calculations have been carried out for the
general model (\ref{0.1}) in original variables, with the use of background
field method and some calculational improvements basically introduced
in similar $d=2$ theory. Our calculational method does not need the
conformal transformation of the metric and therefore is applicable (with minor
standard modifications) to
the higher derivative dilaton model which has been  recently formulated  in
\cite{ejos,shja}.
The theory under consideration leads to a very
cumbersome divergences and hence is non-renormalizable in usual sense.
At the same time if the cosmological (or potential) term $C(\phi)$
is lacking then the theory with the fine tuned functions $A(\phi)$ and
$B(\phi)$ is finite on classical equations of motion, that is possesses the
same property as general relativity. If the  potential term $C(\phi)$
is included the theory is renormalizable on shell, and the divergences
can be removed by the renormalization of the dilaton field.
Summing up, we have constructed 9  versions of the model (\ref{0.1})
with
$$
A(\phi)=a\phi^m,\; \;\;\;\;\;
B(\phi)=b\phi^{m+2}   \; \;\;\;\;\;
C(\phi)=L \phi^k
$$
with $\frac{a}{b}, m, k$ defined in (\ref{3.43}) - (\ref{3.43}),
(\ref{3.111}) - (\ref{3.131}). All these versions have qualitatively the same
renormalization property as Einstein gravity with cosmological term.
They are non-renormalizable off shell and renormalizable on shell. If
the matter fields are included, then the on shell renormalizability is lost.
The higher loops are expected to violate the on shell renormalizability
because of appearance of the counterterms with third powers of curvature.
The one loop  renormalizability of the theory enable us to
apply the renormalization group method for it's study. It turns out that
the effective potential of the theory indicate to the possibility of the
second order phase transitions and in the point of minima the potential
has the form of the Hilbert - Einstein action with both Newtonian and
cosmological constants depending on scale. It is important that the
results of our analysis are independent on the choice of gauge fixing
condition because we consider the on shell renormalization (see \cite{VLT}
for the most complete investigation of the gauge dependence in quantum field
theory.

The one-loop calculations in the model (\ref{0.1}) have been
recently published in \cite{BKK}. In this paper, by use of transformation
like (\ref{1.2}) (the special form of these transformations had been
originally introduced in \cite{hove} for this
purposes) the general model is
reduced to the special case with $A=-\frac{1}{2}, B=const$,
and then the divergences are calculated in a special variables
which correspond to this reduced model.
We have performed the  calculations in an original variables, and in
this sense our result differs from the one of \cite{BKK}.
 In particular, the use of original variables allows the direct application
to the model (\ref{1.12}).
Next, if considered off shell, our counterterms differ
from the ones, derived
in \cite{BKK} because of different choice of quantum variables.
This difference indicate to the parametrization dependence of all the
counterterms. As a consequence, the generalized beta functions, which have
been derived in \cite{BKK}, are likely to be  parametrization
(and probably gauge) dependent. This fact is a direct consequence of
the non-renormalizability of the theory in standard sense.

The calculation in original variables is especially important for the
study of $R+R^2$-gravity. On quantum level this model is equivalent to
some version of (\ref{0.1}), but only in original field variables, since
in the last case one can avoid the introduction of the external source for the
auxiliary scalar field. Introducing an extra constraint on the background
metric, one can derive the renormalization group beta functions
and explore the asymptotics of the effective charges \cite{OV}.

\vspace{5mm}

\noindent{\large \bf Acknowledgments}
The authors are grateful to M. Asorey, I.L. Buchbinder,
T. Morozumi, T. Muta and S.Odintsov for the stimulating discussions.
ILS especially appreciate the contribution of B. Ovrut who paid his
attention to the link between the model (\ref{0.1}) and
special version of higher derivative gravity (\ref{1.12}).
ILS is also grateful to T. Morozumi, T. Muta and to whole Department
of Particle Physics for
warm hospitality during his stay in Hiroshima University,
and the Department of Theoretical Physics at the University of
Zaragoza for warm hospitality at present time.
The  work of ILS has been
supported  in part by the RFFR (Russia), project
no. 94-02-03234, and by ISF (Soros Foundation), grant RI1000.


\newpage

\begin{thebibliography}{99}

\bibitem{DEF} T. Damour and G. Esposito-Far\'ese, Class. Quantum Grav.
{\bf 9}, 2093 (1992); T. Damour and  A. M. Polykov, gr-qc/9411069;
To appear in Gen.Rel.Grav. (1994).

\bibitem{GSW} M.B. Green, J.H. Schwarz and E. Witten, {\it Superstring Theory}
 (Cambridge University Press, Cambridge, 1987).

\bibitem{bar2} J.D. Barrow, Nucl.Phys., {\bf B296}, (1988) 697;
J.D. Barrow, S. Cotsakis, Phys.Lett. {\bf 214B}, (1988) 515;
A.B. Burd, J.D. Barrow, Nucl.Phys., {\bf B308}, (1988) 929.

\bibitem{maeda} K. Maeda, Phys.Rev. {\bf D37}, (1988) 858;
J.D.  Barrow and K. Maeda, Nucl.Phys. {\bf B341}, (1990) 294;
J.D. Barrow, Phys.Rev. {\bf D47}, (1993) 5329;
Phys.Rev. {\bf D48}, (1993) 3592.

\bibitem{wein}  E.J. Weinberg, Phys.Rev. {\bf D40}, (1989) 3950.

\bibitem{CO}  G.L. Cardoso  and B.A. Ovrut,
{\sl Natural Supergravity Inflation}
CERN-TH.6685/92, UPR-0526T (1992); B.A.  Ovrut, {\sl Talk given on
International Conference on Gravity and Field Theory}, Tomsk, August 1994.

\bibitem{DN} T. Damour and K. Nordtvedt, Phys. Rev. {\bf D48}, 3436
(1993).

\bibitem{whit} B. Whitt, Phys.Lett. {\bf 145B}, (1984) 176.

\bibitem{maso} G. Magnano and L.M. Sokolowski,
Phys. Rev. {\bf 50D}, 5039 (1994).

\bibitem{star} A.A. Starobinskii, Phys.Lett. {\bf 91B}, (1980) 99;
L.A. Kofman, A.D. Linde  and  A.A. Starobinskii,  Phys.Lett.
 {\bf 157B}, (1985) 361.

\bibitem{baot} Barrow J.D. and Ottewill A.C., J.Phys. {\bf A16}, (1983) 2757;
Barrow J.D., Phys.Lett., {\bf 183B}, (1987) 285.

\bibitem{MMS} M.B. Mijic, M.M. Morris  and W.-M. Suen, Phys.Rev.
{\bf D34}, (1986) 2934.

\bibitem{hans}  A.A. Starobinskii and  H.J. Schmidt, Class.Quant.Grav.
{\bf 4}, (1987) 695.

\bibitem{9} K.S. Stelle, Phys.Rev. {\bf 16D}, 953 (1977).

\bibitem{vortyu} B.L. Voronov and I.V. Tyutin, Sov. J. Nucl. Phys.
              {\bf 23}, 664 (1976).

\bibitem{11} E.S. Fradkin and  A.A. Tseytlin, Nucl. Phys. {\bf 201B},
469 (1982).

\bibitem{12} I.G. Avramidi, Yad. Fiz. (Sov. J. Nucl. Phys.) {\bf 44},
255 (1986).

\bibitem{bksvw} I.L. Buchbinder and I.L. Shapiro, Yad. Fiz. (Sov. J. Nucl.
Phys.) {\bf 44}, 1033 (1986);
I.L. Buchbinder, O.K. Kalashnikov, I.L. Shapiro,
V.B. Vologodsky and Yu.Yu. Wolfengaut, Phys.  Lett.  {\bf B216}, 127
(1989); I.L. Shapiro, Class. Quant. Grav. {\bf 6}, 1197 (1989).

\bibitem{sh94} I.L. Shapiro, Yad. Fiz. (Sov. J. Nucl. Phys.) (1994)

\bibitem{cosm} S. Christensen and M. Duff., Nucl. Phys. {\bf 170B} (1980) 480.

\bibitem{DW} B.S. DeWitt, {\sl Dynamical Theory of Groups and Fields}.
 (Gordon and Breach, NY, 1965).

\bibitem{book} I.L. Buchbinder, S.D. Odintsov and I.L. Shapiro,
 {\sl Effective Action in Quantum Gravity} (IOP, Bristol, 1992).

\bibitem{hove}  G. t'Hooft and M. Veltman, Ann. Inst. H. Poincare. {\bf A20},
69 (1974).

\bibitem{OV} Paper in preparation.

\bibitem{ejos} E. Elizalde,  A.G. Jacksenaev, S.D. Odintsov and I.L.
 Shapiro,  Phys. Lett. {\bf B328}, 297 (1994);
Preprint HUPD - 9413,
Hiroshima University, 1994; E. Elizalde, S.D.
Odintsov and I.L. Shapiro, Class. Quant. Grav. {\bf 11}, 1607 (1994).

\bibitem{antmot} I. Antoniadis and E. Mottola, Phys. Rev.
      {\bf 45D}, 2013 (1992)

\bibitem{odsh} S.D. Odintsov and I.L. Shapiro, Class. Quant. Grav.
      {\bf 8} L57 (1991).

\bibitem{bosh} I.L. Buchbinder and S.D. Odintsov, Class. Quant. Grav.
      {\bf 2}, 721 (1985).

\bibitem{odsh1} S.D. Odintsov and I.L. Shapiro, Class. Quant. Grav.
      {\bf 9} 873 (1992).

\bibitem{shja} I.L. Shapiro and A.G. Jacksenaev, Phys.Lett. {\bf 324B},
     284 (1994).

\bibitem{25} T. Aida, Y. Kitazawa, H. Kawai and M. Ninomiya,
{\sl Nucl. Phys.} {\bf B427} (1994)  158.

\bibitem{merc} T. Goldman, J. P\'erez-Mercader, F. Cooper, M. Martin-Nieto,
{\sl Phys. Lett.} {\bf281B}  (1992)  219.

\bibitem{lam} I.L. Shapiro, Phys.Lett. {\bf 329B}, 181 (1994).

\bibitem{dene} S.Deser and P. van Nieuwenhuisen,  Phys. Rev.
      {\bf 10D}, 401 (1974); {\bf 10D}, 411 (1974).

\bibitem{vilk} G.A. Vilkovisky, Nucl.Phys. {\bf 234B} 125 (1984).

\bibitem{VLT} Voronov B.L., Lavrov P.M., Tyutin I.V.,
         Sov.J.Nucl.Phys. {\bf 36} 498 (1992).

\bibitem{BKK}  A.O.Barvinski, A.Kamenschik, B.Karmazin,  Phys.Pev. D,
      {\bf 48}, 3677 (1993).

\end{thebibliography}
\end{document}